\documentclass[reprint,preprintnumbers,nofootinbib,amsmath,amssymb,aps,superscriptaddress,prd]{revtex4-2}

\usepackage{amsmath,amssymb}
\usepackage{graphicx}
\usepackage{hyperref}
\graphicspath{{img/}}

\newcommand{\abs}[1]{\left\lvert{#1}\right\rvert}
\newcommand{\absl}[1]{\lvert{#1}\rvert}

\newcommand{\diff}{\mathrm{d}}

\newcommand{\Q}{\ensuremath{{\mathcal{Q}}}}

\begin{document}

\title{Challenging the Stability of Light Millicharged Dark Matter}

\author{Joerg Jaeckel}
\affiliation{Institut f\"ur theoretische Physik, Universit\"at Heidelberg, Philosophenweg 16, 69120 Heidelberg, Germany}

\author{Sebastian Schenk}
\affiliation{Institute for Particle Physics Phenomenology, Department of Physics, Durham University, Durham DH1 3LE, United Kingdom}

\preprint{IPPP/20/81}

\date{May 5, 2021}

\begin{abstract}
We investigate the cosmological stability of light bosonic dark matter carrying a tiny electric charge.
In the wave-like regime of high occupation numbers, annihilation into gauge bosons can be drastically enhanced by parametric resonance. 
The millicharged particle can either be minimally coupled to photons or its electromagnetic interaction can be mediated via kinetic mixing with a massless hidden photon.
In the case of a direct coupling current observational constraints on the millicharge are stronger than those arising from parametric resonance.
For the (theoretically preferred) case of kinetic mixing large regions of parameter space are affected by the parametric resonance leading at least to a fragmentation of the dark matter field if not its outright destruction.
\end{abstract}

\maketitle

\section{Introduction}
\label{sec:introduction}

Very light (sub-eV) dark matter (DM) must consist of bosonic particles. Common examples are axions, axion-like particles or dark photons~\cite{Preskill:1982cy,Abbott:1982af,Dine:1982ah,Nelson:2011sf,Arias:2012az}.
By virtue of their tiny couplings and very low masses, it is often taken for granted that they make good and cosmologically stable candidates for DM.
However, the fact that very light bosonic DM is long lived is far from trivial.
Due to their low mass and low velocity, DM made from light bosons features very high occupation numbers which can dramatically enhance interaction rates with other particles and lead to parametric resonance phenomena~\cite{Kofman:1994rk,Shtanov:1994ce,Kofman:1997yn,Berges:2002cz}.
For example, in significant parts of the parameter space the stability of axion-like particles towards their decay into photons requires a non-trivial interplay of the expansion of the Universe as well as plasma effects~\cite{Abbott:1982af,Preskill:1982cy,Alonso-Alvarez:2019ssa} (cf.~\cite{Kephart:1986vc,Tkachev:1986tr,Tkachev:1987cd,Tkachev:1987ci,Kephart:1994uy,Espriu:2011vj,Tkachev:2014dpa,Rosa:2017ury,Yoshida:2017ehj,Caputo:2018ljp,Caputo:2018vmy,Hertzberg:2018zte,Sawyer:2018ehf,Arza:2018dcy,Sen:2018cjt,Ikeda:2018nhb,Sigl:2019pmj,Arza:2019nta,Wang:2020zur,Arza:2020eik} for some situations where Bose enhancement from high occupation numbers may lead to interesting signatures for axion-like particles).
Of course, these particles do not carry a conserved charge that would naturally render them stable towards decay.
One may therefore wonder what happens if the light DM particles are charged and only annihilations with suitable antiparticles are possible.

To be concrete, in this work, we want to address the question of cosmological ``stability'' for the case that DM carries a tiny electromagnetic charge~\cite{Goldberg:1986nk,Mohapatra:1990vq,Kors:2005uz,Feldman:2007wj,Cheung:2007ut}, often called \emph{millicharge}.
Ample motivation for millicharged particles is provided by Standard Model extensions and, in particular, string theory constructions~\cite{Ignatiev:1978xj,Okun:1983vw,Holdom:1985ag,Dienes:1996zr,Lukas:1999nh,Lust:2003ky,Abel:2003ue,Blumenhagen:2005ga,Batell:2005wa,Blumenhagen:2006ux,Abel:2006qt,Abel:2008ai,Bruemmer:2009ky,Goodsell:2009xc,Shiu:2013wxa}.
At the same time, such scenarios may also offer interesting new opportunities for direct detection~\cite{Berlin:2019uco} (for an overview of various detection strategies, see~\cite{Battaglieri:2017aum}).

In general, having a conserved electric charge, stability towards particle decay is ensured.
However, DM should not carry any net electric charge\footnote{In a scenario where very light bosonic DM does carry an approximately conserved global charge, however, today's net charge density may be non-vanishing~\cite{Alonso-Alvarez:2019pfe}.}.
Therefore, it should be composed of an equal number of particles and antiparticles, opening up the possibility of annihilations, for example into radiation\footnote{In principle, a spatial separation of positive and negative charges may be possible~\cite{Gasenzer:2011by}. While we do not have a conclusive argument excluding this possibility, we strongly suspect that such a situation is not viable in the case where the particles in question carry a gauge charge and are supposed to be the dominant form of DM (e.g.~the presence of long range gauge interactions between the regions may modify the equation of state). Moreover, we note that this is, by definition, connected to a very inhomogeneous situation.}.
Naively, this seems to be strongly suppressed by the tiny value of the relevant charge that is required by phenomenology (for a review see, e.g.,~\cite{Essig:2013lka}), independent of the question of cosmological stability.
However, for low masses, enhancements due to high occupation numbers may set in.
Indeed, in this work, we argue that the coherent nature of the very light DM particles can drastically enhance the interaction rates with gauge bosons.
This, in turn, can cause an annihilation into photons, even for tiny electromagnetic charges.

The basic reason for such an annihilation is that the DM coupling acts as an oscillating mass term for the photons in the coherent DM background.
This can drive the gauge bosons into a parametric resonance~\cite{Kofman:1994rk,Shtanov:1994ce,Kofman:1997yn,Berges:2002cz}, such that certain momentum modes are excited quite rapidly.
The enhancement of the momentum modes then corresponds to an explosive production of photons.
This phenomenon can lead to an efficient depletion of the DM energy density, that seriously challenges the cosmological stability of the DM candidate\footnote{The amplification of gauge fields in the expanding Universe by a parametric resonance from charged scalars has also been considered as a source of large-scale primordial magnetic fields~\cite{Finelli:2000sh}.}.

This work is structured as follows.
In Section~\ref{sec:paramresonance} we discuss the phenomenon of rapid photon production in a millicharged DM background via a parametric resonance.
Furthermore, we carefully examine plasma effects in the early Universe that are able to stop the depletion of the DM energy density.
In Section~\ref{sec:kineticmixing} we consider the theoretically preferred situation where the millicharge arises from a hidden (or dark) photon kinetically mixed with its electromagnetic counterpart.
A brief summary and discussion can be found in Section~\ref{sec:conclusions}.

\section{Resonant Depletion of Millicharged Dark Matter}
\label{sec:paramresonance}

In a scenario where DM carries electromagnetic charge, it is subject to annihilating into visible particles and, in particular, photons.
In this section, we discuss that this depletion of the DM energy density can be drastically enhanced by parametric resonance phenomena~\cite{Kofman:1994rk,Shtanov:1994ce,Kofman:1997yn,Berges:2002cz}, even for tiny charges.
This implies that it is necessary to reconsider the cosmological stability of a millicharged DM candidate.

Let us illustrate the underlying mechanism in a simple setup where a scalar DM candidate is minimally coupled to the photon,
\begin{equation}
	\mathcal{L} = - \frac{1}{4} F^2 + \left( D_{\mu} \phi \right)^{\dagger} D^{\mu} \phi - m^2 \phi^{\dagger} \phi \, .
\label{eq:LagrangianScalarQED}
\end{equation}
Here, $F_{\mu \nu}$ is the electromagnetic field strength associated to the photon $A_\mu$, $\phi$ is the DM of mass $m$ and charge $q$ and $D_{\mu} = \partial_{\mu} + i q A_{\mu}$ denotes the gauge-covariant derivative\footnote{Here, we absorbed the gauge coupling into the definition of the charge, as there is only a single field involved.}.
We are interested in very light, possibly sub-eV, DM particles, which, due to their extremely high occupation numbers, can be described by classical fields.
They furthermore require a non-thermal production which is, for instance, provided by the misalignment mechanism~\cite{Preskill:1982cy,Abbott:1982af,Dine:1982ah,Sikivie:2006ni,Nelson:2011sf,Arias:2012az} (but other mechanisms such as, e.g.,~\cite{AlonsoAlvarez:2019cgw,Ringwald:2015dsf,Co:2017mop,Agrawal:2018vin,Dror:2018pdh,Co:2018lka,Bastero-Gil:2018uel,Long:2019lwl,Peebles:1999fz,Graham:2015rva,Nurmi:2015ema,Kainulainen:2016vzv,Bertolami:2016ywc,Cosme:2018nly,Alonso-Alvarez:2018tus,Markkanen:2018gcw,Graham:2018jyp,Guth:2018hsa,Ho:2019ayl,Tenkanen:2019aij} may also give suitable DM densities).

In first approximation the observed DM energy density can be understood as coherent oscillations of a spatially homogeneous\footnote{Indeed, in the case of the misalignment mechanism and if the field already exists during inflation, any fluctuation is stretched out by inflation, thereby providing for homogeneity.} complex scalar field $\phi$ in the expanding Universe.
We can decompose the field, 
\begin{equation}
    \phi = \varphi \exp \left( i \chi \right) / \sqrt{2} \, .
\end{equation}
In these coordinates, charge neutrality is ensured by trivial dynamics for the angular degree of freedom\footnote{During inflation any non-trivial dynamics of $\chi$ is diluted quickly, $\dot{\chi} \propto a^{-3}$.}, $\chi = \mathrm{const}$, and without loss of generality we can take $\chi=0$.
The radial mode then oscillates with decreasing amplitude
\begin{equation}
\begin{split}
	\varphi(t)  & = \varphi_0 \left( \frac{a_0}{a(t)} \right)^{\frac{3}{2}} \cos \left( m \left(t-t_0\right)\right) \\
	 & = \Phi(t) \cos \left( m \left(t-t_0\right)\right) \, .
\end{split}
\label{eq:oscillation}
\end{equation}
Here, $a(t)$ is the scale factor and $t_0$ denotes the time today with scale factor $a_0$.
Moreover,
\begin{equation}
  \varphi_{0}=\frac{\sqrt{2\rho_{0}}}{m} = 4.5 \times 10^{-6} \, {\rm eV}\left(\frac{{\rm eV}}{m}\right)\left(\frac{\rho_{0}}{\rho_{\mathrm{DM}}}\right)^{\frac{1}{2}}
\end{equation}
is the (average) oscillation amplitude today where we use $\rho_{\mathrm{DM}} = 1.3 \, \mathrm{keV/cm^3}$~\cite{Aghanim:2018eyx}
as a reference value.
The energy density associated to the scalar field then dilutes as $\rho_\phi \sim a^{-3}$, as appropriate for a cold DM particle.

\subsection{Rapid photon production via parametric resonance}

Obviously, the requirement that the energy density of $\phi$ scales like that of pressureless matter is not entirely sufficient for making it the DM.
In addition, a viable DM candidate also has to be cosmologically stable.
Crucially, since in our scenario $\phi$ carries electromagnetic charge, annihilation channels to photons are open, eventually challenging its stability.
In the simple theory~\eqref{eq:LagrangianScalarQED}, the main example of a depletion mechanism would be the pairwise annihilation of DM particles, $\phi \phi \to AA$.
As we will now show, even for tiny electromagnetic charges, the interaction rates of this channel can be significantly enhanced by resonance effects that lead to an explosive production of photons.
To see this, let us consider photon modes in the classical DM background during the evolution of the Universe.
These satisfy
\begin{equation}
	\ddot{A} + H \dot{A} + \left( \frac{k^2}{a^2} + q^2 \varphi^2 \right) A = 0 \, ,
\label{eq:EomPhotons}
\end{equation}
where $A$ denotes a polarization mode of momentum $k$ and $H$ is the Hubble parameter.
$A$ collectively describes the spatial components of the gauge potential $A_{\mu}$, while we fix the temporal components by the Lorentz gauge condition $\partial_\mu \left( \sqrt{-g} A^\mu \right) = 0$.
Once the DM field has overcome the Hubble friction, $H \lesssim m$, it oscillates according to Eq.~\eqref{eq:oscillation} with amplitude $\Phi(t)$.
In this scenario, the equation of motion~\eqref{eq:EomPhotons} can be rewritten as a differential equation of Mathieu type~\cite{McLachlan:1951},
\begin{equation}
	\frac{\diff^2}{\diff x^2} A + \left( \mathcal{A}_k - 2 \mathcal{Q} \cos \left( 2 x\right) \right) A = 0 \, ,
\label{eq:Mathieu}
\end{equation}
with $x = mt$ and we have defined
\begin{equation}
\label{eq:qa}   
	\mathcal{A}_k = \frac{k^2}{a^2 m^2} + \frac{3}{4} \frac{H^2}{m^2} + 2 \mathcal{Q} \, , \enspace \mathcal{Q} = \frac{q^2 \Phi^2}{4m^2} \, .
\end{equation}
The interaction with the millicharged DM acts as an oscillating mass term for the photons.
It is therefore possible that some mode functions are enhanced by resonance effects, known as parametric resonance~\cite{Kofman:1994rk,Shtanov:1994ce,Kofman:1997yn,Berges:2002cz}.
As the mode functions determine the occupation number of the gauge bosons,
\begin{equation}
	n_k = \frac{\omega_k}{2} \left( \frac{\absl{\dot{A}}^2}{\omega_k^2} + \abs{A}^2 \right) \, ,
\end{equation}
this process corresponds to a resonant production of photons.

Crucially, the solution of~\eqref{eq:Mathieu} contains an exponential factor, $A \propto \exp \left( \mu_k x \right)$, with Floquet exponent $\mu_k$.
This exponent is in general a complex number which, importantly, can have a positive real part\footnote{This can, for instance, be read off from the instability chart of $\mathcal{A}_k$ and $\mathcal{Q}$ of the Mathieu equation (see, e.g.,~\cite{McLachlan:1951}).}.
For the purpose of our work, we will exclusively focus on the case where $\mu_k$ is purely real and positive.
This corresponds to an exponential growth of the respective momentum modes,
\begin{equation}
	n_k \propto \exp \left( 2 \mu_k m t \right) \, .
\label{eq:numberdensityexponential}
\end{equation}
That is, the rate of photon production is governed by the Floquet exponent $\mu_k$, which, in general, is a function of $\mathcal{A}_k$ and $\mathcal{Q}$.
Depending on the dynamics of the Mathieu equation, a parametric resonance can be considered in two different regimes.
In a narrow resonance ($\mathcal{Q} \ll 1$) only very few momentum modes are enhanced, while the opposite is true in a broad resonance ($\mathcal{Q} \gg 1$).
Both regimes feature resonant instabilities, which ultimately lead to an explosive production of photons.
For simplicity, let us collect a few important properties of these resonance bands for our discussion.
For details on the dynamics of the Mathieu equation in general and its instability bands in particular we refer the reader to~\cite{McLachlan:1951}.

\subsubsection{Narrow vs broad resonance}

Let us first understand the characteristic behaviour of the Mathieu equation, while neglecting the expansion of the Universe.
This serves as the basis on top of which we can later include the consequences of expansion.

\bigskip

In the narrow resonance regime, where $\mathcal{Q} \ll 1$, the instability bands of the Mathieu equation feature a small width.
In our case, this means that only a limited range of momentum modes is resonantly enhanced.
The first instability band is the dominant one, as it contributes to the exponential growth with the largest Floquet exponent.
The latter is given by~\cite{McLachlan:1951}
\begin{equation}
	\mu_k = \frac{1}{2} \sqrt{ \Q^2 - \left( \frac{k^2}{m^2} - 1 + 2 \Q \right)^2} \, .
\label{eq:narrowmu}
\end{equation}
Furthermore, to good approximation, in momentum space the resonance bandwidth reads~\cite{McLachlan:1951}
\begin{equation}
\label{eq:narrowdelta}
	\Delta k \sim m \mathcal{Q} \, .
\end{equation}
In combination with the requirement $\Q\ll 1$ this justifies the name \emph{narrow} resonance.
Up to corrections of order $\Q$, the Floquet exponent is maximal for momenta of the order of the DM mass, $k_\ast \simeq m$, such that $\mu_{k_\ast} = \mathcal{Q}/2$.
This is essentially the same result that one would obtain from a perturbative approach to interaction rates in $\phi \phi \to AA$ processes, if Bose enhancement is taken into account (see, e.g.,~\cite{Baumann:GraduateCourse}).

\bigskip

In the broad resonance regime, where $\mathcal{Q} \gg 1$, the situation is more complicated.
Here, exact expressions for the Floquet exponents are not available.
Therefore, we will use analytic approximations of $\mu_k$ given in~\cite{Fujisaki:1995ua,Fujisaki:1995dy}.
As their precise form is not very enlightening, we do not quote the full expressions here.
Instead, we give some approximate numbers and behaviours to facilitate the discussion.
That is, typical values of the exponent for a wide range of momenta are $\mu_k \sim 0.15$ and it can obtain a maximum value of $\mu_{k_{\ast}} \sim \log \left(1 + \sqrt{2} \right) / \pi \approx 0.28$~\cite{Fujisaki:1995ua,Fujisaki:1995dy}.
As a rough approximation one can therefore estimate~\cite{Fujisaki:1995ua,Fujisaki:1995dy}
\begin{equation}
\label{eq:broadmu}
    \mu_k \sim 0.15-0.28 \, ,
\end{equation}
within the resonance bandwidth.
Importantly, this does not strongly depend on $\Q$.
The width of the instability band is then typically of the order~\cite{Kofman:1997yn}
\begin{equation}
\label{eq:broaddelta}
	\Delta k \sim m \mathcal{Q}^{\frac{1}{4}} \, ,
\end{equation}
which can be parametrically large for $\mathcal{Q} \gg 1$.

Moreover, we note that in the broad resonance regime we typically have to take multiple instability bands into account, once the expansion of the Universe is considered.
Luckily, the above rough expressions hold for all of them.
In addition, we remark that the typical distance in $\Q$ between resonance bands of the exponent $\mu_k \left( \mathcal{A}_k(\Q), \Q \right)$, for momenta of the order $k \lesssim \Delta k$, is
\begin{equation}
    \Delta \Q \sim \sqrt{\Q} \, .
\end{equation}

The narrow and broad resonance regime can behave very differently with respect to the depletion of the DM energy density.
Generally speaking, a broad resonance is typically more efficient, as more momentum modes are within the resonance band at the same time and the exponent is typically larger.
That said, for our discussion the dependence of the rate of exponential growth, given by $\mu_k$, as well as the width of the resonance $\Delta k$ on the parameter $\Q$ is important.
In the narrow resonance regime, the former is parametrically given by $\mu_k \sim \mathcal{Q}$ while for a broad resonance it is typically of the order $\mu_k \sim 0.15$, largely independent of $\Q$.
The bandwidth of both regimes is also considerably modified, i.e.~$\Delta k \sim m \mathcal{Q}$ for a narrow and $\Delta k \sim m \mathcal{Q}^{1/4}$ for a broad resonance, respectively.
As we will see in the following section, both aspects are crucial for an explosive production of photons.

\subsubsection{Including the expansion of the Universe}

The previous discussion of the dynamics of the Mathieu equation only applies to a static situation.
However, in the early Universe the expansion cannot be neglected.
In this case, the parameters $\mathcal{A}_k$ and $\mathcal{Q}$ explicitly depend on the scale factor.
Strictly speaking, the concept of (static) resonance bands then ceases to be meaningful.
Nevertheless, if the changes are sufficiently slow (compared to the time-scale of the DM oscillations) we can still get a reasonable picture by imagining the movement of a given momentum mode $k$ along the trajectory $\left(\mathcal{A}_{k}(t) , \, \Q(t) \right)$ through the instability chart.
For instance, the system might start to evolve within a broad resonance, but as $\mathcal{Q}$ decreases with time, it eventually ends up in a narrow resonance regime before it terminates.

In a situation where $\mathcal{Q}$ may change significantly between consecutive oscillations of the driving DM field $\phi$, one would instead have to move from a parametric resonance to a so called \emph{stochastic} resonance~\cite{Kofman:1997yn}.
Here, with a single oscillation of $\phi$ the phase of each photon mode is drastically altered, such that they are practically uncorrelated at any stage of photon production.
Therefore, due to interferences, the number of photons produced typically increases but can also decrease with progressing DM oscillations, thereby slightly reducing the efficiency of the resonance.
For a detailed discussion of the phenomenon of a stochastic resonance in an expanding Universe, we refer the reader to~\cite{Kofman:1997yn}.

Such a thorough treatment of stochastic resonance is beyond the scope of this work.
We therefore follow the more intuitive approximate approach outlined above, which has already been pursued in~\cite{Alonso-Alvarez:2019ssa}.
For our purpose, the most important difference between the static and the dynamical situation is that the photons experience a redshift as the Universe expands.
Mathematically, the Mathieu parameter $\mathcal{A}_{k}$ directly depends on the physical momentum of the mode, $ \propto k/a$, which changes with time.
Each mode therefore only spends a finite amount of time in the resonant region.
This can prevent the DM from efficiently annihilating into photons, if the latter are shifted out of a resonance quickly enough.

After a short amount of time $\delta t$, the momentum of a photon mode is shifted by~\cite{Alonso-Alvarez:2019ssa}
\begin{equation}
	\frac{\delta k}{k} \simeq H \delta t \, ,
\label{eq:redshift}
\end{equation}
where $\delta t$ is thought to be differential on cosmological scales, $\delta t \ll H^{-1}$.
That is, the momentum modes of the photons can only grow exponentially for a short amount of time, $\delta t_{\exp} \sim 1 / (2 \mu_k m)$, before they get shifted out of the resonance band, $\delta t_{\exp} \lesssim \delta t$.
While this is a universal feature that eventually terminates the resonant production of gauge bosons, its physical manifestation within the Mathieu dynamics has to be established carefully.
In particular, there can be differences between a narrow and a broad resonance due to the significant modifications of the instability chart in these regions.
Hence, we will discuss both scenarios separately.

\bigskip

\paragraph{Narrow resonance}

In a narrow resonance the Floquet instabilities occur at integer values of $k/m$ (see also Fig.~\ref{fig:mdm-instabilitychart}).
In this regime, the lowest instability band, corresponding to the momentum $k_\ast \simeq m$, is the most effective and we focus on this in our analysis.
Hence, effectively, there is only a single resonance band that can induce an exponential growth of the photon modes.
At the same time the expansion of the Universe can redshift the photons out of this instability, thereby preventing their resonant enhancement.
That is, naively, there is a competition between the characteristic time of exponential growth determined by the Floquet exponent and the time the photon modes spend inside the resonance band.
Reversing this argument, it can be written as a naive condition to \emph{avoid} the rapid production of photons in the DM background,
\begin{equation}
	\frac{1}{2 \mu_{k_\ast} m} \gtrsim \frac{\delta k}{k_\ast} \frac{1}{H} \, .
\label{eq:StabilityNaive}
\end{equation}
Obviously, this requirement is time dependent.
Loosely speaking, the condition has to be satisfied at all times in the narrow resonance regime in order to avoid the complete fragmentation of the DM field and thereby to guarantee the cosmological stability of the DM candidate.

The stability condition~\eqref{eq:StabilityNaive} so far only takes into account the growing exponential factor of the photon mode functions.
However, as there can still be a small prefactor in front, a single short burst of rapid photon production may not be sufficient to trigger a complete annihilation of the DM field.
Instead, the latter is only effective, if a significant amount of energy is transferred from the DM to the photons, i.e.~if $\rho_A / \rho_\phi$ grows sufficiently that it becomes of order unity\footnote{When $\rho_A / \rho_\phi \sim 1$, we expect our description of the coherent DM field to break down and backreaction effects to become important. We will comment more on this later. For now, we note that this is the reason we often use the word ``fragmentation'' indicating the loss of the coherent condensate instead of speaking of ``annihilation''.}.
This, in turn, can be used to obtain a more precise condition for its cosmological stability.
While the DM energy density schematically reads $\rho_\phi \simeq m^2 \Phi^2 / 2$, the energy density of the photons can be obtained by summing over all modes\footnote{Here, we consider one polarization mode. In principle, the two polarizations of the photons grow equally fast leading to a factor of 2 in the energy density, which, however, has only a negligible effect on the limits we will derive.}
\begin{equation}
	\rho_A = \frac{1}{\left(2 \pi a\right)^3} \int \diff^3 k \, \omega_k n_k \, ,
\label{eq:EnergyDensityPhotons}
\end{equation}
where $\omega_k$ denotes the energy of each momentum mode.
Indeed, if the sub-exponential prefactor of $n_k$ in~\eqref{eq:numberdensityexponential} is small, the resonance needs to be active for a considerable amount of time to transfer a significant amount of energy from the DM field into photons, 
\begin{equation}
	\delta T = \frac{\zeta}{2 \mu_k m} \, .
\end{equation}
In practice, the factor $\zeta$ depends on the initial conditions associated to the photon mode functions.
Following~\cite{Alonso-Alvarez:2019ssa}, the sub-exponential correction can be estimated via a saddle-point approximation of~\eqref{eq:EnergyDensityPhotons}.
It schematically reads
\begin{equation}
	\zeta \sim \log \left( \frac{1}{n_0} \sqrt{\frac{m}{H}} \frac{\Phi}{qm} \right) \, ,
\end{equation}
where $n_0$ is the initial occupation number of the photon modes, which can be determined by vacuum fluctuations or CMB photons, $n_0 = 1/2$ or $n_0 \simeq 2 T_{\mathrm{CMB}} / m$, respectively.
The prefactor $\zeta$ can then be chosen conservatively, i.e.~corresponding to the larger of both options.

Finally, requiring that the photons are redshifted out of the resonance quickly enough to avoid a complete fragmentation of the DM field yields a condition for the stability of the DM candidate,
\begin{equation}
	\zeta \gtrsim 2\frac{\Delta k}{k_{\ast}} \frac{m}{H}\mu_{k_{\ast}} \, .
\label{eq:DMstabilitycondition}
\end{equation}
Using $\Delta k / k_{\ast} = \Q$ and $\mu_{k_{\ast}} = \Q / 2$, this condition reads $\zeta \gtrsim (m/H) \Q^2$ (see also~\cite{Kofman:1997yn}).
Therefore, in a narrow resonance, to avoid a fragmentation of the DM field its electromagnetic charge $q$ has to satisfy
\begin{equation}
	\log \left( \frac{1}{n_0} \sqrt{\frac{m}{H}} \frac{\Phi}{qm} \right) \gtrsim \frac{m}{H} \left( \frac{q \Phi}{2m} \right)^4 \, .
\label{eq:DMstabilityconditionNarrow}
\end{equation}
In principle, to guarantee stability this inequality must hold at all times of the cosmic evolution.
However, to ensure consistency, the system has to be in the narrow resonance regime, $\Q \ll 1$. In general, this is not necessarily the case.
For example, as typically $\zeta\sim 10-100$, this condition is not consistent with the narrow resonance regime for $m/H \sim \mathcal{O} (1)$~\cite{Kofman:1997yn}.
It is therefore worthwhile to also consider the broad resonance regime.

\bigskip

\paragraph{Broad resonance}

In a broad resonance the instability chart of the Mathieu equation is more complicated.
In contrast to the case of a narrow resonance, the instability bands are not sharply localized around integer values of $k/m$.
Instead, for a fixed $\Q$, they can extend from a typical scale of $k_\ast \sim m \Q^{1/4}$ down to possibly even vanishing momentum (see also Fig.~\ref{fig:mdm-instabilitychart}).
This means that, as the Universe expands, a given momentum within a certain instability band is not redshifted out of a \emph{single} resonance, but, because $\Q$ decreases similarly as $\Q \sim a^{-3}$, it can cross multiple instability bands before it enters the regime of a narrow resonance.
Therefore, the photon mode can experience multiple resonant enhancements on its trajectory through the instability chart.
As an approximation we can model this by summing up all resonances that a given $k$-mode crosses,
\begin{equation}
	n_k \propto \exp \left(2 m \int \diff t \, \mu_k(t) \right) \, .
\end{equation}
Since the Mathieu parameter $\Q$ is now a function of time, in a radiation-dominated Universe the exponent can be written as
\begin{equation}
	m \int \diff t \, \mu_k(t) = \frac{1}{3} \frac{m}{H} \left(\frac{q \Phi}{2m} \right)^{\frac{4}{3}} \int \diff \Q \, \frac{\mu_k \left( \Q \right)}{\Q^{\frac{5}{3}}} \, .
\end{equation}
As pointed out above, within the instability bands for the range of momenta $k \leq \Delta k \sim m \Q^{1/4}$, the Floquet exponent takes typical values of $\mu_k \sim 0.15 - 0.28$, which are mostly independent of $\Q$ to good approximation.
In this case, the bands can be assumed to be of width and distance of order $\sqrt{\Q}$ in $\Q$.
Hence, the integration of $\mu_k$ is dominated by small values of $\Q$.
As a rough approximation we can therefore choose the integration boundaries to be $\Q_{-} = 1$ (beginning of the broad resonance region) and $\Q_{+} = \infty$.
Evaluating this numerically we find
\begin{equation}
	\kappa (k) = \int_1^{\infty} \diff \Q \, \frac{\mu_k \left( \Q \right)}{\Q^{\frac{5}{3}}} \approx
	\begin{cases}
    	0.13	\, , & k = 0 \\
		0.046 \, , & k = m \Q^{1/4}
	\end{cases} \, ,
	\label{eq:kappa}
\end{equation}
with monotonically decreasing values within those limiting cases.

Similar to the case of a narrow resonance, the fragmentation of the DM field is efficient (ultimately leading to a breakdown of our description of the DM oscillations), if a significant fraction of energy is transferred from the DM to the photons, $\rho_A / \rho_\Phi \sim 1$.
Again, this requirement can be translated into a stability condition for the DM candidate,
\begin{equation}
	\log \left( \frac{8 \pi^{3/2}}{\sqrt{3} n_0} \sqrt{\frac{m}{H}} \frac{\kappa}{q^2 \epsilon^3} \right) \gtrsim \frac{2\kappa}{3} \frac{m}{H} \left( \frac{q \Phi}{2m} \right)^{\frac{4}{3}} \, ,
\label{eq:DMstabilityconditionBroad}
\end{equation}
where we take account of modes up to $k=\epsilon m \Q^{1/4}$ in the energy density.
This is the broad resonance equivalent to the stability requirement~\eqref{eq:DMstabilityconditionNarrow}, which is applicable in the narrow resonance regime.

Note that in determining this expression we have made several rough approximations.
As already stated above we have neglected the dependence on the upper integration boundary in~\eqref{eq:kappa}.
This ignores contributions suppressed by an inverse power of $\Q$.
We have also used that, due to the exponential growth in the modes, the biggest drain in energy occurs at late times and therefore evaluated the energy drain only at the end of the broad resonance regime.
Moreover, using a fixed $\kappa$ we have neglected that during the evolution the physical momentum of each mode decreases as $k \sim a^{-1}$ and therefore the exponent changes.
Finally, in line with all these approximations we have simply dropped terms logarithmic in $\Q$.

\bigskip

\paragraph{Discussion}

The stability conditions~\eqref{eq:DMstabilityconditionNarrow} and~\eqref{eq:DMstabilityconditionBroad} put strong constraints on the value of the electromagnetic charge of the DM.
Before explicitly evaluating them, let us first get some analytical understanding.
In principle, in order to avoid the resonant depletion of the DM, both conditions have to complement each other such that either one of them is satisfied at all times of the cosmic evolution.
As we have pointed out before, we expect the fragmentation of the DM field to first be governed by a broad resonance.
Then, as $\Q$ decreases with time in an expanding Universe, $\Q \sim a^{-3}$, the system will enter the narrow resonance regime before the fragmentation eventually terminates (given that the coherent field is not completely destroyed at that point).
Therefore, in practice, one has to carefully establish which stability condition gives the correct, i.e.~self-consistent, constraint on the millicharge at each time.
This depends on the charge as well as on the time when the stability condition is evaluated.

\begin{figure}
	\centering
	\includegraphics[width=0.8\columnwidth]{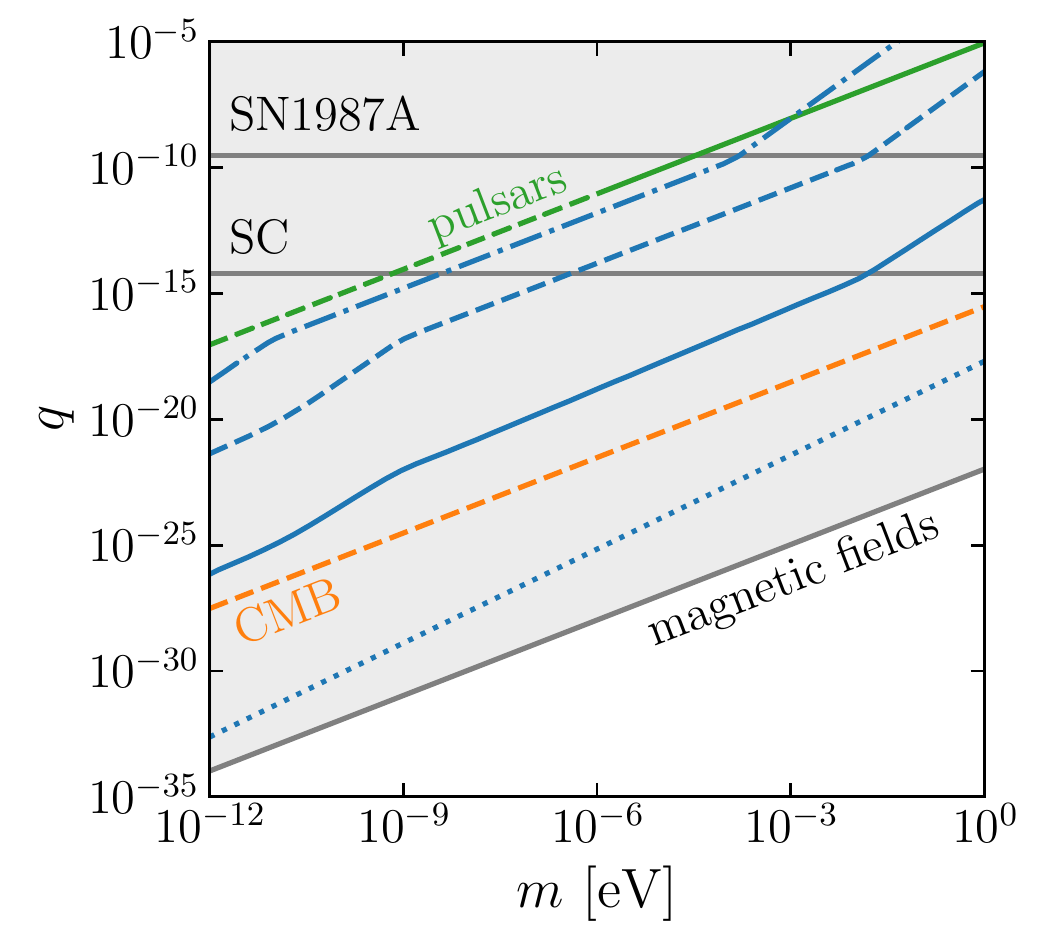}
	\caption{Allowed electromagnetic charge $q$ of the scalar DM candidate as a function of its mass $m$. The solid blue line corresponds to the stability condition due to parametric resonance, evaluated at the time where plasma effects terminate the latter, i.e.~where $m_A \simeq m$. The dashed and dash-dotted blue lines illustrate an estimate of where plasma photons with $m_{A}>m$ may be produced due to a broad resonance, Eq.~\eqref{eq:ResonanceHighMass}, evaluated when $m_A / m = 1$ and $m_A / m = 100$, respectively. For comparison, the dotted blue line shows the narrow resonance condition evaluated close to the earliest possible time, $1000 \, t_\ast$, while plasma effects are neglected, $m_A = 0$. The observational constraints are given by CMB observations~\cite{Xu:2018efh}, SN1987A~\cite{Chang:2018rso} and stellar cooling (SC)~\cite{Davidson:2000hf}, pulsar timing arrays~\cite{Caputo:2019tms} or by interactions with magnetic fields in galaxies and clusters~\cite{Kadota:2016tqq,Stebbins:2019xjr}. In these limits, the dashing indicates regions where we have used very naive extrapolations from the high-mass regime.}
\label{fig:mdm-paramspace_plasma}
\end{figure}

In general, it is \emph{a priori} not at all obvious, what time gives the strongest possible constraint on $q$.
In fact, both regimes~\eqref{eq:DMstabilityconditionNarrow} and~\eqref{eq:DMstabilityconditionBroad} behave very differently with the scale factor.
To see this, we insert the evolution of the DM amplitude, $\Phi \sim a^{-3/2}$, and the behaviour of the Hubble constant during radiation domination, $H \sim a^{-2}$, into the exponential factor (i.e.~the right hand side) of both stability conditions.
For a narrow resonance we obtain
\begin{equation}
	\frac{m}{H} \left( \frac{q \Phi}{2m} \right)^4 \sim a^{-4} \, ,
\label{eq:scalingnarrow}
\end{equation}
while, in contrast, the broad resonance regime behaves as
\begin{equation}
	\frac{2\kappa}{3} \frac{m}{H} \left( \frac{q \Phi}{2m} \right)^{\frac{4}{3}} \sim \mathrm{const} \, .
\label{eq:scalingbroad}
\end{equation}
This suggests that in the narrow resonance regime the strongest constraint arises when evaluating at the earliest possible time, whereas the broad resonance case appears to be independent of the scale factor.

Let us consider both scenarios.
In case of a narrow resonance, Eq.~\eqref{eq:scalingnarrow} suggests that we should evaluate the stability condition when $\phi$ just starts to oscillate, i.e.~at $t_{\ast}$ when $H \simeq m$.
However, as noted before, fulfilling~\eqref{eq:DMstabilityconditionNarrow} with $\zeta \sim 10-100$ is inconsistent with the narrow resonance regime at $t_\ast$.
Hence, we either have to evaluate at a somewhat later time when $H\ll m$ or go into the regime of a broad resonance.
For the strongest self-consistent constraint on the millicharge in the narrow resonance regime we can evaluate at $t \approx 1000 \, t_\ast$.

In contrast, in the broad resonance regime, Eq.~\eqref{eq:scalingbroad} suggests that the constraint on the millicharge is independent of the time when the stability condition is evaluated.
This suggests that the limit does not strengthen much when approaching the broad resonance regime\footnote{We note that the complete independence of the evaluation time is, of course, due to the simplistic approximations we employ.}.
Indeed, we have checked that~\eqref{eq:DMstabilityconditionBroad} roughly provides for the same limit as~\eqref{eq:DMstabilityconditionNarrow} evaluated at $t \approx 1000 \, t_\ast$.

A numerical evaluation of the stability condition~\eqref{eq:DMstabilityconditionNarrow} at $t_1 \equiv 1000 \, t_\ast$ is shown as the dotted blue line in Fig.~\ref{fig:mdm-paramspace_plasma}.
However, this estimate is probably too optimistic as we still need to include plasma effects, which we will do next.

\subsection{Photons inside the early Universe plasma}
\label{subsec:plasma}

So far, we have assumed that the photons are moving freely through the Universe.
However, during the cosmological evolution, the early Universe is filled with a hot plasma that modifies their propagation.
Indeed, for example in the case of axion-like particles, this is the dominant effect ensuring their stability~\cite{Preskill:1982cy,Abbott:1982af,Alonso-Alvarez:2019ssa}.
Therefore, it is sensible to also consider this effect for the case of millicharged DM (a discussion of parametric resonance in charged cosmological scalars can also be found in~\cite{Finelli:2000sh} but they were focused on primordial magnetic fields rather than DM).

Naively, the photons interact with the charged particles of the medium such that they acquire a modified dispersion relation (and wavefunction renormalization), see e.g.~\cite{Raffelt:1996wa}.
Effectively they acquire a mass, $m_A$.
In Eqs.~\eqref{eq:Mathieu} and \eqref{eq:qa} this leads to the replacement $k^2/a^2 \to k^2/a^2 + m_A^2$. 
Therefore, $\mathcal{A}_{k}$ is larger and, if the plasma mass is too high, the resonance becomes inefficient.
In particular, in a narrow resonance regime the instabilities become ineffective if the plasma mass exceeds the mass of the DM candidate, $m_A \gtrsim m$.
For nearly all $k$ the rate of exponential growth $\mu_k$ becomes imaginary, corresponding to an oscillating rather than a growing mode function.
In contrast, in a broad resonance, the production of photons with masses $m_A \gg m$ is in principle possible\footnote{We thank Paola Arias, Ariel Arza and Diego Vargas for very useful discussions (triggered by the helpful comments of an anonymous referee for~\cite{Arias:2020tzl}) on this issue in a similar system.}.
However, this process requires comparatively large couplings in general.
In particular, photon production in this regime can only be efficient for charges satisfying~\cite{Kofman:1997yn}
\begin{equation}
	q \gtrsim 4 \frac{m_A^2}{m \Phi} \, .
\label{eq:ResonanceHighMass}
\end{equation}
Overall, we therefore expect that this possibility will result in a weaker constraint on the electromagnetic millicharge (see also the example below).

The plasma mass of the photon depends on the temperature of the medium.
That is, in an expanding Universe, it is time dependent\footnote{We use the cosmological evolution of $m_A$ as given in~\cite{Alonso-Alvarez:2019ssa} which is based on~\cite{Raffelt:1996wa,Braaten:1993jw,Redondo:2008ec,Mirizzi:2009iz,Dvorkin:2019zdi}.}.
As noted above, in practice, the condition $m_A \lesssim m$ sets the earliest time at which the (narrow resonance) stability condition~\eqref{eq:DMstabilityconditionNarrow} can be evaluated and turns out to stabilize the scalar DM candidate in large parts of parameter space of the vanilla theory of millicharged DM.
We show this as a solid blue line in Fig.~\ref{fig:mdm-paramspace_plasma}.

In the broad resonance regime, the production of photons with masses above the DM mass is possible, but for this, larger charges are needed.
In particular, as a minimal requirement, the broad resonance must be strong enough to overcome the mass threshold. This requires fulfilling the condition~\eqref{eq:ResonanceHighMass}.
This is usually already a weaker requirement than evaluating the narrow resonance stability condition at the point at which the photon mass is small enough for the narrow resonance to be active.
As an example, this is demonstrated by a dashed and a dashdotted blue line in Fig.~\ref{fig:mdm-paramspace_plasma}, where we choose times when the plasma mass is of the order of $m_A / m = 1$ and $m_A / m = 100$, respectively.

Looking at Fig.~\ref{fig:mdm-paramspace_plasma} we can see the drastic impact of the plasma effects.
Comparing the naive estimate that completely neglects plasma effects (dotted blue line) with the constraint taking into account the plasma effects (solid blue line), the former turns out to be many orders of magnitude stronger.
Indeed, we observe that in the regime of very low masses, the stability condition on millicharged DM is a weaker requirement than current observational constraints~\cite{Xu:2018efh,Chang:2018rso,Davidson:2000hf,Caputo:2019tms,Kadota:2016tqq,Stebbins:2019xjr}\footnote{As indicated also in the figure, for some constraints we have extremely naively extrapolated to very small masses. Moreover, we note, that most of the DM constraints have been derived having at least implicitly particles in mind. It may therefore be worthwhile to rethink and check their validity in the fully wave-like regime. In this sense the stability constraints may even find some non-trivial application in this model. In this case the caveats on coherence and backreaction discussed in the next section should, however, also be taken into account.}.
While this is desirable from a physical point of view, the simple model we have considered in this section is disfavoured from a theoretical point of view, as quantization of electromagnetic charge would be hard to justify.
Let us therefore turn to a more realistic and appealing theory involving a hidden photon that is kinetically mixed with the visible sector.

\section{Millicharged Particles arising from a Kinetic Mixing of a Hidden Photon with the Visible Sector}
\label{sec:kineticmixing}

From a theoretical point of view, a fundamental millicharge is unappealing with respect to charge quantization.
A well motivated alternative is provided by kinetic mixing~\cite{Holdom:1985ag}.
A simple example of this scenario is an additional massless hidden photon $X_{\mu}$ that is kinetically mixed with the electromagnetic photon $A_{\mu}$~\cite{Holdom:1985ag},
\begin{equation}
	\mathcal{L}_\epsilon = -\frac{\epsilon}{2} F_{\mu \nu} X^{\mu \nu} \, .
\end{equation}
For a hidden sector matter particle $\phi$ (in our case the DM candidate) carrying a (quantized) charge $q$ under $X_{\mu}$, a small effective electromagnetic charge appears after diagonalizing the kinetic term.
To see this, one can rotate the gauge fields by $A_{\mu} \to A_{\mu}$ and $X_{\mu} \to X_{\mu} + \epsilon A_{\mu}$.
While this redefinition leads to canonically normalized kinetic terms of the gauge fields, it also appears in the gauge-covariant derivatives of the hidden sector matter field,
\begin{equation}
    D_{\mu}\phi=(\partial_{\mu}+qg X_{\mu})\phi \rightarrow (\partial_{\mu}+qg X_{\mu}+\epsilon q g A_{\mu}) \phi \, ,
\end{equation}
where $g$ is the hidden sector gauge coupling.
As a consequence, the DM carries an effective electromagnetic charge~\cite{Holdom:1985ag}
\begin{equation}
	q_{\mathrm{eff}} = \epsilon q g  \, ,
\label{eq:chargekineticmixing}
\end{equation}
where we have again absorbed the factor of the electromagnetic coupling $e$ into the charge.
In this way, a small $\epsilon$ (and possibly also $g$) can lead to a tiny electromagnetic charge, even if $q$ is integer\footnote{Such a situation has, for instance, been explored in order to mediate long-range forces between hidden sector particles (see, e.g.,~\cite{Ackerman:mha,Foot:2014uba}).}.

In general, the amount of kinetic mixing is a free parameter and $\epsilon$ may even be of order one.
However, if we consider the hidden photon to be part of a hidden sector we usually expect that the mixing is small.
For instance, the hidden gauge group may be understood as a low-energy remnant of a UV theory with a unified gauge symmetry broken at some high scale~\cite{Bruemmer:2009ky}.
After symmetry breaking, some degrees of freedom, for instance a heavy fermion, usually carry a charge both under electromagnetism as well as the hidden gauge group.
Quantum mechanically, a kinetic mixing between both gauge fields is then induced by a fermion loop of the UV theory.
At low energies, the kinetic mixing parameter is determined by the corresponding one-loop Feynman diagram and parametrically reads (see, e.g.,~\cite{Holdom:1985ag,Bruemmer:2009ky})
\begin{equation}
	\epsilon \sim \frac{e g}{6 \pi^2} \log \left( \frac{m_{\psi}}{\mu} \right) \, ,
\label{eq:loopinducedkineticmixing}
\end{equation}
where $m_{\psi}$ is the mass of the heavy fermion and $\mu$ is the regularization scale of the loop integral.
This typically gives a small kinetic mixing, which is particularly tiny if also the hidden sector gauge coupling is small, $g \ll 1$.

\bigskip

We can now apply the arguments of the previous section to this scenario.
As we assume no plasma to be present in the hidden sector, we expect that annihilation into hidden photons remains possible and therefore guaranteeing stability may put stronger constraints on the millicharge.

We will examine this scenario in two separate steps.
First we completely neglect the (small) kinetic mixing effects, i.e.~we consider a secluded hidden sector without kinetic mixing, $\epsilon = 0$.
Then we argue that the main conclusions also hold in the phenomenologically more interesting case with a small but non-vanishing kinetic mixing parameter.

\subsection{Secluded hidden sector}

\begin{figure}
	\centering
	\includegraphics[width=0.925\columnwidth]{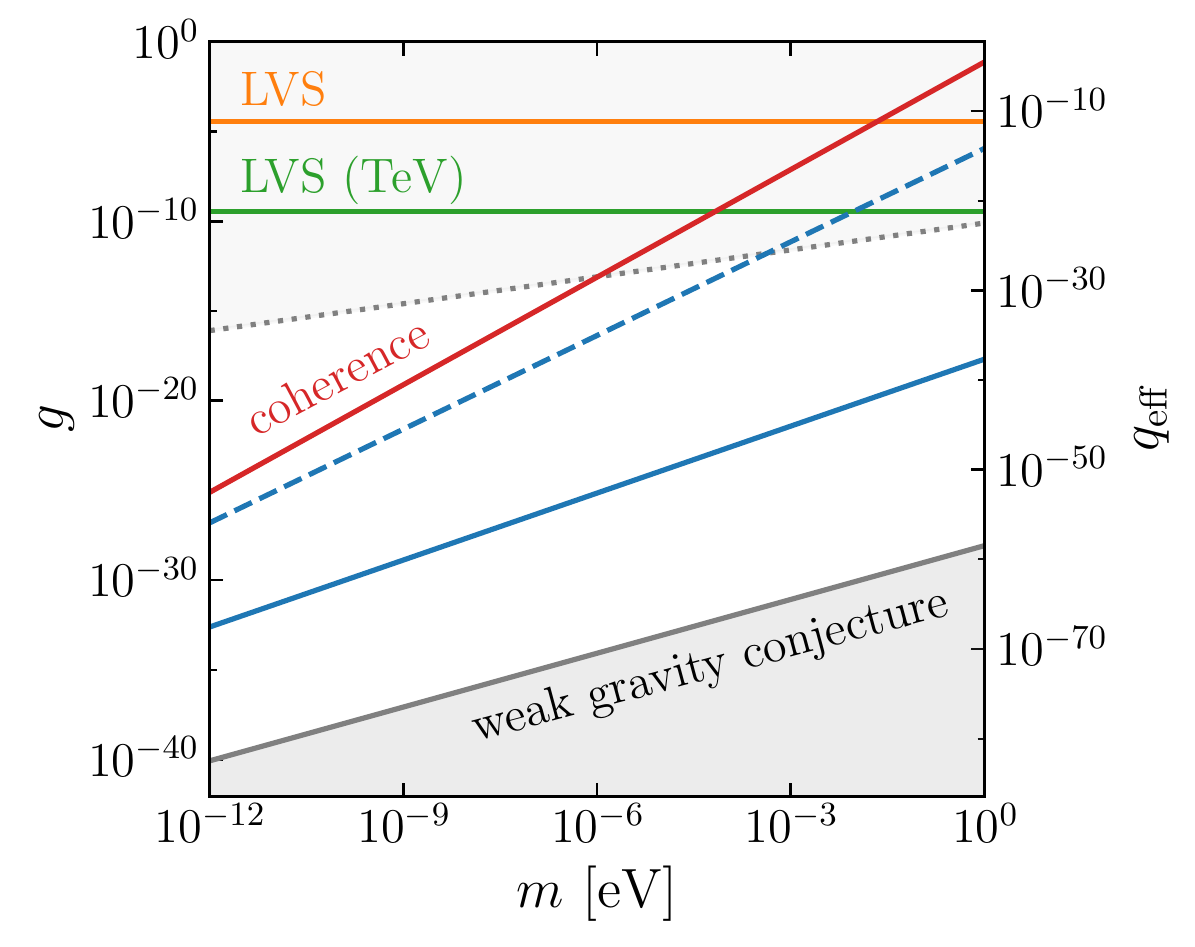}
	\caption{Allowed hidden gauge coupling $g$ to the scalar DM candidate as a function of its mass $m$. The blue lines correspond to the stability condition due to the parametric resonance, evaluated close to the time where the field starts to oscillate, $t_1 \approx 1000 \, t_{\ast}$, (solid) and at matter-radiation equality (dashed). The red line is given by a coherence condition, discussed in the main text. For comparison, the right axis shows the typical corresponding effective millicharge induced by a fermion-loop of a UV theory, $q_{\mathrm{eff}} \sim \epsilon g \sim e g^2 / (6 \pi^2)$. Along the same lines, the light-shaded grey area in the upper region corresponds to observational constraints provided by interactions with a magnetized intergalactic stellar medium~\cite{Kadota:2016tqq,Stebbins:2019xjr}, see Fig.~\ref{fig:mdm-paramspace_plasma}.}
\label{fig:mdm-HP-paramspace}
\end{figure}

In the case $\epsilon=0$, our discussion in Section~\ref{sec:paramresonance} completely carries over.
In particular, the DM stability conditions~\eqref{eq:DMstabilityconditionNarrow} and~\eqref{eq:DMstabilityconditionBroad} in the narrow and the broad resonance regime can be applied at all times of the cosmic evolution.
Most importantly, as there is no effective mass of the hidden photon that could block the resonant enhancements, they can in principle be satisfied at the earliest possible time.
As discussed in the previous Section~\ref{sec:paramresonance}, a reasonable estimate is obtained by evaluating the stability condition from the narrow resonance regime at $t_1 \approx 1000 \, t_{\ast}$.
This is shown as the solid blue line in Fig.~\ref{fig:mdm-HP-paramspace}, where we display the allowed value of $g$ as a function of the DM mass $m$.
(Note that here, we have normalized the field to unit charge, $q=1$.)
The requirement of avoiding a resonant depletion of the DM energy density into hidden photons puts severe constraints on the hidden gauge coupling for small masses.

\subsubsection{Backreaction effects}

The above evaluation might be an overestimation of the constraint posed by the DM stability requirement.
This is because, so far, we have neglected the backreaction of the parametric resonance on the DM field.
Obviously, a first effect is the depletion of the DM field.
This is what we have implicitly used to set our constraint, i.e.~using energy conservation to determine the depletion from the produced gauge bosons.
However, if the energy density in the hidden photons is comparable to that in the DM field, $\rho_A \sim \rho_\phi$, it is conceivable that energy starts to be transferred back to the DM field, slowing down the depletion.
Although this is non-trivial due to the fact that most of the produced hidden photons have momenta $k \simeq m$, processes involving multiple hidden photons may be possible due to the high occupation numbers and the resonantly enhanced interaction rates.
With the present analysis we cannot exclude this possibility. 
A thorough analysis of this effect would need to involve some careful numerical simulations, which is beyond the scope of this work.
The overall allowed value of the hidden gauge coupling might therefore be higher.

That said, let us obtain a very conservative estimate of the point where the resonance should shut off.
Allowing for the backreaction effect, we nevertheless expect that in such a situation a significant fraction of the total energy in the DM-hidden photon system, possibly $\sim 1/2$, will be in hidden photons and therefore in the form of dark radiation.
At matter-radiation equality such a large fraction of dark radiation is certainly excluded (cf., e.g.~\cite{Akrami:2018vks}).
Therefore, we can evaluate the DM stability condition at matter-radiation equality.
At this stage, at the latest, $\phi$ is required to behave as standard cold DM.
At the same time, we expect the system to be in a narrow resonance regime, such that the stability condition~\eqref{eq:DMstabilityconditionNarrow} is valid.
The resulting constraint on the hidden gauge coupling is shown as a dashed blue line in Fig.~\ref{fig:mdm-HP-paramspace}.
It is considerably weaker than the original estimate, but still affects an appreciable region of parameter space, bearing in mind that this estimate is probably overly conservative.

\subsubsection{Non-trivial initial momentum distribution}

In addition to the fragmentation of the light scalar DM field via parametric resonance, there is another physical effect that may modify the interaction rate between the DM and the hidden photons.
Crucially, in our analysis we treat $\phi$ as a spatially homogeneous classical field.
While such a situation arises naturally in the misalignment~\cite{Preskill:1982cy,Abbott:1982af,Dine:1982ah,Sikivie:2006ni,Nelson:2011sf,Arias:2012az} effect when the field is extremely homogenized by inflation, other production mechanisms (see, e.g.,~\cite{AlonsoAlvarez:2019cgw,Ringwald:2015dsf,Co:2017mop,Agrawal:2018vin,Dror:2018pdh,Co:2018lka,Bastero-Gil:2018uel,Long:2019lwl,Peebles:1999fz,Graham:2015rva,Nurmi:2015ema,Kainulainen:2016vzv,Bertolami:2016ywc,Cosme:2018nly,Alonso-Alvarez:2018tus,Markkanen:2018gcw,Graham:2018jyp,Guth:2018hsa,Ho:2019ayl,Tenkanen:2019aij}) typically feature a non-trivial momentum distribution\footnote{Alternatively, this could also be due to the backreaction effect, which likely produces DM particles of non-vanishing momentum.} for the millicharged particles.
Hence, in such a scenario, the field exhibits spatial variations.
This can be approximately taken into account by ensuring that there is a sufficient amount of coherence.

This has been discussed in detail in~\cite{Arias:2020tzl} (see also~\cite{Tkachev:2014dpa,Hertzberg:2018zte,Arza:2018dcy,Wang:2020zur,Arza:2020eik} for discussions in the context of DM structures) from which we summarize the main implications.
In order to preserve coherence of the hidden photons produced by the resonance, the width of the resonance in momentum space has to be larger than the momentum spread of the DM, $\Delta k \gtrsim \Delta k_{\phi}$.
The latter can be estimated to be $\Delta k_{\phi} \sim m v_{mr} \left(a_{mr} / a\right)$, where we require that $\phi$ should be non-relativistic at matter-radiation equality, $v_{mr} \sim 10^{-3}$ (cf., e.g.~\cite{Colombi:1995ze,Bode:2000gq,Cooray:2003dv,Kunz:2016yqy}).
Therefore, the condition for preserving coherence can be written as
\begin{equation}
	\Delta k \gtrsim m v_{mr} \left( \frac{a_{mr}}{a} \right) \, .
\label{eq:coherence}
\end{equation}
As pointed out in Section~\ref{sec:paramresonance}, the width of the resonance bands in momentum space, i.e.~the left-hand side of this inequality, depends on the value of the Mathieu parameter $\mathcal{Q}$.
Evaluating~\eqref{eq:coherence} at matter-radiation equality we see that the required width is much smaller than the mass $m$.
Therefore, we can use the narrow resonance regime where $\Delta k \sim m \mathcal{Q}$.
This can be immediately translated into a constraint on the hidden gauge coupling, which we similarly evaluate at matter-radiation equality.
This is shown in red in Fig.~\ref{fig:mdm-HP-paramspace}.

\subsubsection{Discussion}

In general, our results, shown in Fig.~\ref{fig:mdm-HP-paramspace}, demonstrate that the stability requirement for a very light DM candidate in a secluded hidden sector affects sizeable regions of parameter space.
The strongest constraint is posed by the parametric resonance stability condition evaluated close to the time, when the DM field starts to oscillate (solid blue).
However, this neglects backreaction effects and therefore needs to be taken with caution.
A more conservative estimate is given by evaluating the stability condition at matter-radiation equality (dashed blue).
We expect that a more careful numerical analysis would most likely reveal a stability condition that lies in between those possibilities.
Aside from backreaction effects, additionally, a non-trivial initial velocity distribution of the DM particles, possible in some models for their production, may allow to weaken the constraint as the resonance requires a sufficient amount of coherence.
Taking into account that the DM velocity must be small enough to allow for successful structure formation, we find the most conservative constraint shown as the red line.
Intriguingly, we are still able to probe large regions of parameter space.
In fact, it seems challenging to motivate or construct models featuring gauge groups with such tiny gauge couplings.
A famous example allowing for small gauge couplings is provided by the large volume scenario (LVS) of type IIB string compactifications~\cite{Balasubramanian:2005zx,Conlon:2005ki}.
Here, the gauge theory is supported on D-branes wrapping cycles of the internal Calabi-Yau manifold of ten-dimensional spacetime.
The volume of these internal cycles, in turn, determines the gauge coupling, $g \sim \mathcal{V}^{-1/3}$~\cite{Burgess:2008ri}.
Therefore, hyper-weakly coupled gauge theories can be engineered by choosing an appropriate Calabi-Yau geometry that supports large D-brane worldvolumes.
In Fig.~\ref{fig:mdm-HP-paramspace}, we show typical values of the gauge coupling achieved in a generic (orange) and low string-scale, $M_S \sim 1 \, \mathrm{TeV}$, (green) LVS~\cite{Burgess:2008ri}.
Strikingly, even the extreme case of TeV-scale strings is rendered unstable for a wide range of masses even using the most conservative constraint.
We note, however that the bound provided by the weak gravity conjecture~\cite{ArkaniHamed:2006dz}, shown in grey, is not reached. It would therefore be interesting to see whether consistent models with such small gauge couplings can be constructed.

As we will argue in the following section, similar conclusions still hold, if the hidden sector is not completely secluded but has a small kinetic mixing parameter connecting it to the visible world.

\subsection{Non-vanishing kinetic mixing}

Along the lines discussed at the beginning of the section we focus on a situation with small kinetic mixing.
In a homogeneous background $\varphi$, the equations of motion are linear in the photon $A$ and hidden photon $X$ and therefore couple distinct momentum modes of both fields.
In fact, after having redefined the gauge fields by $A_{\mu} \to A_{\mu}$ and $X_{\mu} \to X_{\mu} + \epsilon A_{\mu}$, their mode functions satisfy
\begin{equation}
\begin{split}
	\ddot{A} + H \dot{A}  + \left( \frac{k^2}{a^2} + m_A^2 +  \epsilon^2 g^2 \varphi^2 \right) A &= \epsilon g^2 \varphi^2 X \, , \\
	\ddot{X} + H \dot{X}  + \left( \frac{k^2}{a^2} + g^2 \varphi^2 \right) X &= \epsilon g^2 \varphi^2 A \, .
\end{split}
\label{eq:eomkineticmixing}
\end{equation}
Here, we have already included an effective mass for the photon, $m_A$.
Naively, the equations of motion imply that both the photon as well as the hidden photon modes can be enhanced by a parametric resonance induced by the oscillating DM background.
However, a resonant enhancement of $A$ is now parametrically weaker as compared to $X$, because its coupling contains an additional factor of the mixing parameter $\epsilon$.
This means that, for instance, modes of the hidden photons might be growing rapidly due to a broad resonance, while the photon modes already are in a very narrow resonance regime and not amplified efficiently.
At the same time, this amplification might also act as an oscillating driving force on the right hand side of~\eqref{eq:eomkineticmixing}.
Eventually, the growing modes of both fields will converge to the same resonance frequency after a certain period of time.
Therefore, in general, the DM may largely annihilate into hidden photons and also, to a smaller fraction, into visible photons which follow shortly after.

\begin{figure}
	\centering
	\includegraphics[width=0.8\columnwidth]{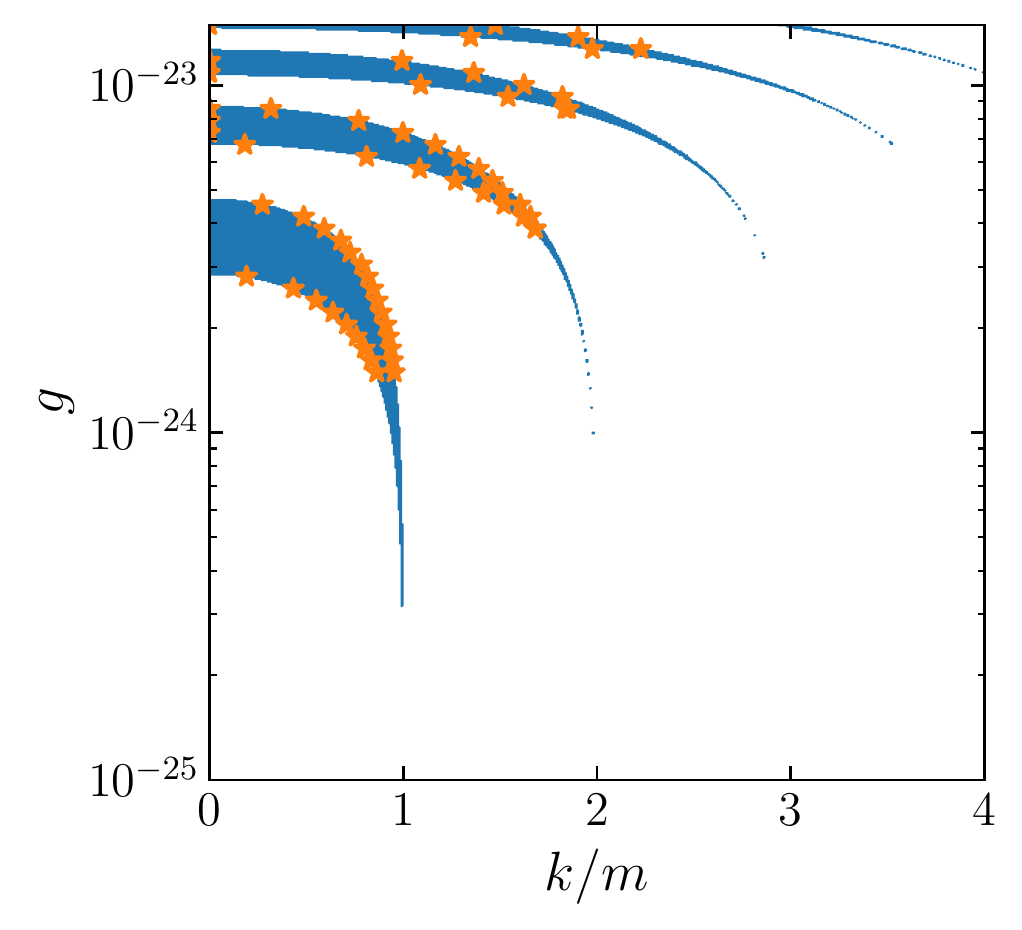}
	\caption{Floquet instabilities of the Mathieu equation for $X$ with $m = 10^{-3} \, \mathrm{eV}$ evaluated at $t_{\ast}$ (blue). The orange dots illustrate an approximation of the same instabilities for the kinetic mixing case with $\epsilon = 0.1$ and $m_A / m = 100$. Inside the blue bands, and in between the orange points, the mode functions can grow exponentially.}
\label{fig:mdm-instabilitychart}
\end{figure}

The above observations suggest that an efficient depletion of the DM energy density is possible in a theory featuring kinetic mixing.
In practice, this is important, as the visible photon can obtain a non-negligible plasma mass, $m_A \neq 0$, while the hidden photon is still massless. 
However, as the DM mainly annihilates into hidden photons, we are able to avoid plasma effects of the visible photons in the early Universe almost entirely.
This is illustrated in Fig.~\ref{fig:mdm-instabilitychart} where we compare the instability chart of a completely secluded hidden photon (blue) with that for non-vanishing kinetic mixing, $\epsilon=0.1$ and $m_A /  m = 100$ (orange points denoting the boundary).
We obtain these by numerically solving the coupled equations of motion for $X$ and $A$ for different momenta.
As an example, we choose a DM mass of $m = 10^{-3} \, \mathrm{eV}$ and both instabilities are evaluated when the DM field starts to oscillate, $t_{\ast}$.
Inside the instability bands, an exponential growth of the momentum modes of $X$, i.e.~rapid production of hidden photons, is possible.
We can see that the unstable regions are almost identical.
The plasma mass in the visible sector does not prevent the resonant annihilation of DM into hidden photons.
We expect this to be true everywhere in parameter space for kinetic mixing parameters smaller\footnote{We have also checked examples with large kinetic mixing, $\epsilon \sim 1$. In this case, the exponential growth may be absent.} than $\epsilon \lesssim 0.1$.
Therefore, the allowed values of the hidden gauge coupling are almost identical to what is shown in Fig.~\ref{fig:mdm-HP-paramspace}.
To give an impression of the constraints of the effective millicharge, we show on the right-hand axis of Fig.~\ref{fig:mdm-HP-paramspace} indicative values of the millicharge obtained by combining Eqs.~\eqref{eq:chargekineticmixing} and \eqref{eq:loopinducedkineticmixing}.
The light grey region indicates the experimental and observational constraints on the effective charge as also shown in Fig.~\ref{fig:mdm-paramspace_plasma}.
There are large regions where even our most conservative estimate of the unstable region poses a stronger constraint on the effective millicharge than current observational bounds.

\section{Conclusions}
\label{sec:conclusions}

The microscopic nature of dark matter (DM) that comprises large parts of the cosmic fabric remains elusive.
As suggested by its name, so far, there is no experimental evidence of DM interacting with electromagnetism.
While this naively rules out any sizable electric charge assigned to DM particles, it is still possible that their charge is tiny, thereby strongly suppressing interactions with photons.
In this work, we have investigated the cosmological longevity of such DM particles in the sub-eV mass regime.
In this mass range the DM particles must be bosonic and, for concreteness, we have chosen them to be scalar.

The millicharged particles are either minimally coupled to photons or their electromagnetic interaction is mediated via kinetic mixing with a massless hidden photon.
In both cases, due to the large occupation numbers of the light DM field, even for tiny charges the DM may efficiently annihilate into gauge bosons via a parametric resonance~\cite{Kofman:1994rk,Shtanov:1994ce,Kofman:1997yn,Berges:2002cz}.

We find that, in the case of a direct coupling to photons current observational constraints on the millicharge are stronger than those arising from parametric resonance, as shown in Fig.~\ref{fig:mdm-paramspace_plasma}.
This is mainly due to the plasma mass that the photons acquire in the hot medium of the early Universe, which essentially terminates the resonance if it is larger than the DM mass.
In contrast, in the case of a theory featuring kinetic mixing, plasma effects are practically absent.
Therefore, even employing conservative estimates large regions of parameter space are affected by the parametric resonance as illustrated in Fig.~\ref{fig:mdm-HP-paramspace}.
In fact, in particular for very low DM masses, its electric millicharge has to be orders of magnitude below what has been typically obtained in UV models, e.g.~in type IIB string compactifications in the large volume limit~\cite{Balasubramanian:2005zx,Conlon:2005ki,Burgess:2008ri} (we note that, indeed, already the experimental and observational constraints rule out this region of parameter space for loop-induced kinetic mixing).
That said, the limits do not yet reach the smallest possible values suggested by the weak gravity conjecture~\cite{ArkaniHamed:2006dz}, which are many orders of magnitude below the smallest values found in the concrete realizations discussed above.

We conclude that it is far from trivial that very light bosonic DM carrying a tiny electric charge is long lived.
Instead, its high occupation numbers can dramatically enhance interaction rates with gauge bosons leading to parametric resonance phenomena.
Therefore, its cosmological stability cannot be taken for granted but has to be considered carefully.

\begin{acknowledgments}
We are grateful to Patrick Foldenauer, Arthur Hebecker, Ruben Kuespert and Michael Spannowsky for valuable discussions.
We also thank Paola Arias, Ariel Arza and Diego Vargas for collaboration on related work.
S.S.~is funded by the Deutsche Forschungsgemeinschaft (DFG, German Research Foundation) -- 444759442.
\end{acknowledgments}

\bibliography{refs}

\end{document}